\documentclass[aps,prc,nofootinbib,amsmath,amssymb, onecolumn]{revtex4-2}
\usepackage{graphicx}% Include figure files
\usepackage{dcolumn}% Align table columns on decimal point
\usepackage{color}

\newcolumntype{P}[1]{>{\centering\arraybackslash}p{#1}}
\begin{document}
\title{Electric field tunable magnetoexcitons in Xenes/hBN/TMDC, Xenes/hBN/BP, and Xenes/hBN/TMTC heterostructures}
\author{Roman Ya. Kezerashvili$^{1,2}$, Anastasia Spiridonova$^{1}$, and Klaus Ziegler$^{1,3*}$}
%\thanks{E-mail contact: klaus.ziegler@physik.uni-augsburg.de}  }
\affiliation{$^{1}$New York City College of Technology, The City University of New York, Brooklyn, NY 11201, USA\\
$^{2}$The Graduate School and University Center, The City University of New
York, New York, NY 10016, USA \\
$^{3}$Institut f\"ur Physik, Universit\"at Augsburg, D-86135 Augsburg, Germany \\
$^*$corresponding author: klaus.ziegler@physik.uni-augsburg.de
}
\date{\today}
%\begin{abstract}
\begin{abstract}
In this work, we propose novel van der Waals (vdW) heterostructures composed of Xenes, transition metal dichalcogenides (TMDCs), phosphorene, and transition metal trichalcogenides (TMTCs),
which are separated by insulating hexagonal boron nitride (hBN) layers. We theoretically investigate the
behavior of Rydberg indirect excitons in Xenes/hBN/TMDC, Xenes/hBN/BP, and Xenes/hBN/TMTC
heterostructures subjected to parallel external electric and magnetic fields that are oriented perpendicular to the layers.
By incorporating both
isotropic and anisotropic materials, we demonstrate that excitonic properties can be effectively tuned through the external field
strengths and the heterostructure design.

Our results show that the exciton reduced mass and the binding energy increase with the electric field strength, while enhanced
dielectric screening from additional hBN layers reduces the binding energy. Anisotropic materials exhibit distinct excitonic responses,
including variations in diamagnetic behavior. Moreover, the diamagnetic energy contributions and coefficients decrease with stronger
electric fields but increase with the number of hBN layers. Finally, we explore the potential of time-periodic electric fields with Floquet
band-structure engineering. These findings provide a comprehensive framework for controlling excitonic phenomena in
low-dimensional materials. %, enabling the design of advanced optoelectronic and quantum devices.

\end{abstract}
\maketitle

\section{Introduction}

Two-dimensional layered materials have attracted persistent attention and have been extensively studied over the past two decades. Graphene has sparked significant research interest in two-dimensional (2D) layered materials such as transition metal dichalcogenides (TMDCs) \cite{Wang2018} and group IVA materials such as silicene, germanene, and stanene, collectively known as Xenes, which consist of atoms arranged in a honeycomb lattice similar to that of graphene, but with varying degrees of buckling of two sublayers with respect to each other \cite{Matthes2013,Molle2017}. In contrast to graphene, these materials exhibit either direct or indirect band gaps and possess remarkable physical properties as a result of their crystal symmetry and reduced dimensionality. Moreover, TMDCs and Xenes typically display in-plane structural isotropy in their electrical and optical properties.

Phosphorene, the monolayer form of black phosphorus (BP), was synthesized in 2014 \cite{Bridgman1914} and offers several advantages over TMDCs and Xenes. Its most notable features include strong in-plane anisotropy, a thickness-dependent band gap, and high carrier mobility. The monolayer structure of BP appears to be composed of two distinct planes forming a puckered honeycomb structure \cite{Carvalho2016,Batmunkh2016,Qiu2017}, which gives rise to its anisotropic electronic characteristics.

A newer class of 2D materials exhibiting strong in-plane anisotropy includes the group IVB transition metal trichalcogenides (TMTCs). These materials also feature a puckered honeycomb structure and have garnered increasing interest in recent years \cite{KezSpir2022tmtc}. Prototypical examples such as TiS$_3$, ZrS$_3$, and ZrSe$_3$ have been successfully synthesized \cite{Patra2020}. TMTCs are composed of atomic layers held together via weak van der Waals forces and exhibit quasi-one-dimensional behavior similar to that of phosphorene. Their unique and highly anisotropic crystal structures lead to drastically different material properties along different crystallographic directions \cite{Rodin2014,Li2014b,Lv2014,Chaves2015,Dai2016,Qiu2017,Le2019,Brunetti2019,Kamban2020,Patra2020,Yoon2021,Kezspir2022superfl}.

Due to the structural differences between the armchair (AC) and zigzag (ZZ) directions, both phosphorene and TMTCs exhibit strong in-plane anisotropy. Many of their physical properties, including the effective masses of charge carriers, vary significantly between these two directions. This anisotropic nature contrasts the in-plane isotropy observed in TMDCs and Xenes. TiSe$_3$, ZrS$_3$, and ZrSe$_3$ monolayers are all indirect-gap semiconductors \cite{Jin2015,Ming2015}, whereas phosphorene is a direct-gap semiconductor with a highly anisotropic dispersion near the band edge.

The concept of exciton formation in systems with spatially separated electrons and holes—specifically in coupled quantum wells where electrons and holes occupy different wells—was first introduced in Ref.~\cite{LozovikYudson}, inspiring a wealth of theoretical and experimental studies on double quantum wells and heterostructures composed of 2D materials. Today, heterostructures made from layers of TMDCs, phosphorene, Xenes, TMTCs, and others are at the forefront of nanomaterials research. The formation of excitonic complexes such as excitons, trions, and biexcitons in these systems leads to unique electronic, optical, and mechanical properties, which differ significantly from their bulk counterparts and hold great promise for applications in nanoelectronics, optoelectronics, and quantum computing \cite{Optoelectronic2021,QuantumCom2020}.

Heterostructures comprising 2D monolayers separated by dielectric layers—particularly when there is a lattice mismatch between the constituent layers, as, for example, in MoS$_2$/hBN/WSe$_2$ or MoSe$_2$/hBN/WSe$_2$—have been extensively studied over the past two decades. These investigations have uncovered rich physics arising from interlayer coupling, moiré superlattices, and excitonic phenomena, providing deep insights into their electronic properties and excitonic behaviors, see  \cite{Thygesen2017,Kamban2020,Jiang2021} and references herein. %Such studies underscore the critical role of lattice mismatch in designing and understanding the properties of 2D heterostructures, paving the way for novel applications in optoelectronics and quantum devices.
Combining different 2D materials into heterostructures enables the tuning of electronic and optical properties  \cite{Yankowitz2019,Xian2021}. % and flat electronic bands \cite{Xian2021}. via a precise control over stacking sequences, twist angles, and interlayer interactions. For example, a relative twist between layers can lead to the formation of moiré patterns that significantly influence the electronic band structure and may give rise to phenomena such as superconductivity \cite{Yankowitz2019} and flat electronic bands \cite{Xian2021}.

The band structures of TMDC, phosphorene, and TMTC monolayers are largely unaffected by a perpendicular electric field, and the same holds true for heterostructures composed of these monolayers. In contrast, Xenes exhibit a tunable band gap due to their buckled structure. Therefore, heterostructures such as Xenes/hBN/TMDC, Xenes/hBN/BP, or Xenes/hBN/TMTC can exhibit a tunable gap, which in turn affects the properties of indirect excitons formed within them. The buckled geometry of Xenes leads to a potential difference between sublattices, allowing control over the band gap, effective masses of charge carriers, binding energies (BEs), and diamagnetic coefficients (DMCs) of magnetoexcitons. We do not consider systems where Xene monolayers are placed directly on top of other layers without hBN separation, as Xenes are unstable in air \cite{Acun2013,Tao2015}.

In 2D systems, the reduced dielectric screening and strong Coulomb interactions enhance the binding of excitons, making them especially sensitive to magnetic fields. When a magnetic field is applied, it affects both the binding energy and the diamagnetic shift of the exciton.
In van der Waals heterostructures, composed of novel 2D materials, the impact of a magnetic field on excitons is further influenced by the nature of interlayer coupling, effective masses, and dielectric environments. These systems allow for tunable magnetoexcitonic behavior through effective mass anisotropy \cite{KezSpir2022phosp,KezSpir2022tmtc}, layer engineering, and electric field control.

In this article, we propose van der Waals (vdW) heterostructures engineered by combining layers of Xenes and TMDCs, as well as Xenes and phosphorene or TMTCs, separated by hBN layers. We investigate the properties of excitons in these heterostructures under the influence of external electric and magnetic fields. The electric and magnetic fields are parallel and oriented perpendicular to the plane of the vdW heterostructure.

The goals of this article are twofold: (i) to demonstrate the tunability of the properties of indirect excitons in Rydberg states within isotropic (Xenes/hBN/TMDC) and anisotropic (Xenes/hBN/BP or Xenes/hBN/TMTC) heterostructures using external electric and magnetic fields; and (ii) to explore the use of a time-periodic electric field as a tool for band-structure engineering.
The remainder of this article is organized as follows. In Section~\ref{theory}, we present the theoretical model used to describe the electron-hole system in the presence of external electric and magnetic fields, both with and without effective mass anisotropy of the charge carriers. Section~\ref{results} contains the results of our calculations for the binding energies of Rydberg exciton states under an external electric field in vdW heterostructures. In the same section, we also present and discuss the BEs and DMCs of magnetoexcitons in these heterostructures under the influence of an external electric field. Additionally, we explore a band-structure engineering via a time-periodic electric field. Finally, our concluding remarks are given in Section~\ref{ConcludRemark}.

\begin{figure}[b]
\begin{centering}
\includegraphics[width=6.3cm]{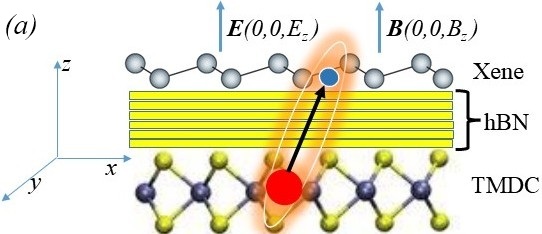}
\includegraphics[width=8.0cm]{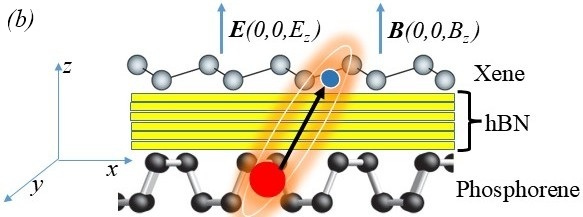}
\caption{(Color online) Schematic illustration of excitons in ($a$) Xenes/hBN/TMDC and ($b$) Xenes/hBN/BP van der Waals heterostructure in electric and magnetic fields. By replacing phosphorene by TMTC monolayer one gets Xenes/hBN/TMTC heterostructure.
}
\label{fig1M}
\end{centering}
\end{figure}

\section{Theoretical Model}
\label{sect:model}

\label{theory}

It is known that electrostatically bound electrons and holes in an external magnetic field form magnetoexcitons. In this section, following Refs.~\cite{Gorkov1967,Spiridonova,KezSpir2021,RKAS2021b}, we briefly introduce the theoretical model for describing Mott-Wannier magnetoexcitons in Xenes/hBN/TMDC and Xenes/hBN/BP or Xenes/hBN/TMTC heterostructures. We consider the energy contribution, $\Delta E$, from external electric and magnetic fields to the binding energies, $E(B)$, of Rydberg states of magnetoexcitons and their diamagnetic coefficients. In the system under consideration, excitons are confined within a van der Waals heterostructure, where $N$ layers of hBN monolayers separate Xenes and TMDC or Xenes and phosphorene or Xenes and TMTC monolayers.

Let us outline the low-energy model that describes exciton states in Xenes/hBN/TMDC and Xenes/hBN/BP van der Waals heterostructures
in the $x-y$ plane. We consider an electron and a hole in parallel magnetic
$\mathbf{B} = (0, 0, B_z) \equiv (0, 0, B)$ and electric $\mathbf{E} = (0, 0, E_z) \equiv (0, 0, E)$ fields,
which are perpendicular to the heterostructure, as depicted in Fig.~\ref{fig1M}.
Monolayers of silicene, germanene, and low-buckled stanene can be pictured as honeycomb graphene monolayers with an out-of-plane buckling, such that the A and B triangular sublattices are offset with respect to the $x-y$ plane with a typical distance $d_{0}$ to the latter. This distance is known as the buckling constant or buckling factor. The intrinsic sensitivity of Xenes
to a perpendicular electric field is due to the offset between the two triangular sublattices.
In particular, the resulting asymmetry  causes an on-site potential difference between the sublattices.
In the absence of an external electric field the band structure of Xenes in the vicinity of the $K/K^{\prime }$ points
resembles gapped graphene, %though, the intrinsic gaps of Xenes are significantly larger than that of graphene.
while the application of a perpendicular electric field changes the band gap in the monolayer
Xenes.Thus, the single-particle dispersion of quasiparticles in a monolayer Xenes
%in the electric field acting along the $z$-axis
is described in the vicinity of the $K/K^{\prime }$ points by
the two-dimensional (2D) Hamiltonian \cite{Tabert2014}
\begin{equation}
\label{dirac1}
H_0=v_F(\xi p_x{\hat\tau}_x+p_y{\hat\tau}_y)-\xi\Delta_{so}
{\hat\sigma}_z{\hat\tau}_z+\Delta_z{\hat\tau}_z
%E(p)=\sqrt{\Delta _{\xi \sigma }^{2}+v_{F}^{2}p^{2}}
,
\end{equation}
where ${\hat\sigma}_j$ and ${\hat\tau}_j$ are Pauli matrices in the spin and
in the pseudospin space, respectively. The pseudospin is a result of the
underlying honeycomb lattice. $\xi=\pm1$ refers to the valley index of the
honeycomb lattice. Moreover, $v_F$ is the Fermi velocity, $\Delta_{so}$ is the
spin-orbit splitting, and $\Delta_z=ed_0E_{\perp}$ describes the potential caused by
a perpendicular electric field $E_{\perp}$ on the two triangular sublattices of the
honeycomb lattice. Since only ${\hat\sigma}_z$ appears, the Hamiltonian
$H_0$ separates in the spin space into two massive 2D Dirac Hamiltonian with dispersions
\begin{equation}
\label{dispersion1}
    E_{\sigma}(p)=\pm\sqrt{\Delta_{\xi\sigma}^2+v_F^2p^2}
\end{equation}
with $\Delta_{\xi\sigma}=|\xi \sigma \Delta _{so}-ed_{0}E_{\perp}|$, where
$\xi, \sigma =\pm 1$ are the valley and spin indices, respectively.
Thus, we have four bands with two different gaps $\Delta _{\xi \sigma }$ due
to $\xi,\sigma=\pm1$.
The shift of $\Delta _{\xi \sigma }$ by the electric field $E_{\perp}$ was previously
discussed in Refs.
\cite{Drummond2012, Ezawa1, Ezawa2, Ezawa3}, as well as in Ref. \cite{Tabert2014}.
For the following, we assume that
the kinetic energy $v_F^2p^2$ is small compared to $\Delta_{\xi\sigma}^2$,
such that we can expand the square root to obtain a parabolic dispersion
$E_\sigma(p)\sim \pm(|\Delta _{\xi \sigma }|+p^2/2m)$
with the effective electron mass $m=|\Delta _{\xi \sigma}|/v_F^2$ \cite{Pan2015,Matthes2013}.
Thus, the conduction and valence bands are parabolic in the vicinity of the
$K/K^{\prime }$ points.  
The effective masses
of electrons and holes are the same due to the symmetry between the lowest
conduction and highest valence bands, and can be written as a function of
the external electric field in the following form:
\begin{equation}
m^{e}=\frac{\lvert \xi \sigma \Delta _{so}-ed_{0}E_{\perp}\rvert }{%
v_{F}^{2}}.  \label{eq:effmassEz}
\end{equation}

At non-zero electric fields,
both the valence and conduction bands, split into upper bands with a large
gap (when $\xi =-\sigma $), and lower bands with a small gap (when $\xi
=\sigma $). We call the excitons formed by charge carriers from the large
gap $A$ excitons, and those formed by charge carriers in the small gap $B$
excitons.
At small electric fields, germanene and especially stanene show significant
differences between the masses of the $A$ and $B$ excitons. The latter
is due to their large intrinsic band gaps. Silicene, which has an intrinsic
band gap on the order of a couple of meV, exhibits very little difference
between the masses of $A$ and $B$ excitons, even at relatively small
electric fields. At large electric field the difference between the $A$ and
$B$ exciton mass is negligible in silicene and germanene. In all
cases, the mass of the $A$ exciton exceeds the mass of the $B$ exciton.
TMDC monolayers are isotropic and effective masses of an electron and hole are different. % and $m_h > m_e$.

In the next step, we will consider a system whose translation invariance is
broken by an external magnetic field. To this end, let us introduce the coordinate vectors of the electron and hole for the
Mott-Wannier exciton in the Xenes and TMDC (phosphorene or TMTC) layers.
The following in-plane
coordinates $\mathbf{r}_{e}(x_{e},y_{e})$ for an electron in Xenes layer and $\mathbf{r}_{h}(x_{h},y_{h})$
for a hole in TMDC or phosphorene layer, respectively, are used in our description.
The electron/hole spins are polarized by the magnetic field, and the honeycomb structure is irrelevant because the system is dominated by the
magnetic length. Then the Hamiltonian of an interacting pair of an electron
at the site $r_e$ and a hole at the site $r_h$ reads ($\hbar=c=1$)
\begin{eqnarray}
\label{hamiltonian2}
\hat{H}=&&\frac{1}{2m^{e}}\Big( i\nabla _{x}^{e}-eA_{x}(r_{e})\Big) ^{2}+%
\frac{1}{2m^{e}}\Big( i\nabla _{y}^{e}-eA_{y}(r_{e})\Big) ^{2}+\frac{1}{%
2m_{x}^{h}}\Big( i\nabla _{x}^{h}+eA_{x}(r_{h})\Big) ^{2}+\frac{1}{%
2m_{y}^{h}}\Big(  i\nabla _{y}^{h}+eA_{y}(r_{h})\Big) ^{2}  \notag \\
&&+V\Big( |\mathbf{r}_{e}-\mathbf{r}_{h}|\Big),  \label{eq:Schredingermag}
\end{eqnarray}
where 
$m_{x}^{h}$ and $m_{y}^{h}$ correspond to the effective mass of the
hole in the $x$ or $y$ direction in the phosphorene or TMTC, respectively.% In the Xenes monolayer $m_{x}^{e}=m_{x}^{h}\equiv m^{e}$ and the Hamiltonian (\ref{eq:Schredingermag}) becomes simpler.

%.The effective mass  In the case of Xenes--TMDC heterostructure $m_{x}^{e}=m_{x}^{h}=m_{h}\equiv m_{e}$ and the Hamiltonian (\ref{eq:Schredingermag}) becomes simpler.
In Eq. (\ref{eq:Schredingermag})
$V\left( |\mathbf{r}_{e}-\mathbf{r_{h}}|\right) $ describes the Coulomb
interaction between the electron and hole via a central Coulomb potential
\begin{equation}
V\left( \sqrt{x^{2}+y^{2}+D^{2}}\right) =-\frac{ke^{2}}{\kappa \left( \sqrt{x^{2}+y^{2}+D^{2}}\right) },  \label{eq:indcoul}
\end{equation}%
where $r^2 = x^{2}+y^{2}$. %The effective masses $m^{e}$ and $m^{h}$ are taken from the calculation of the expanded square-root dispersion of Eq. (\ref{dispersion1}).
The anisotropic nature of the phosphorene and TMTC, in contrast to other 2D isotropic materials such as graphene and TMDC semiconductors,
breaks the circular symmetry. This requires
the use of Cartesian coordinates for the description of excitons. The
asymmetry of the electron and hole dispersion in phosphorene is reflected in the
Hamiltonian for the Mott-Wannier exciton. The Hamiltonian
for an interacting electron-hole pair in Xenes/hBN/BP or Xenes/hBN/TMTC heterostructures in
the external electric and magnetic fields is obtained within the framework of
the effective mass approximation. In particular, the electric field dependence of Hamiltonian (\ref{hamiltonian2})
is implicit via the effective mass of electron in Xenes monolayer.
Equation~
\eqref{eq:indcoul} describes the interaction between the electron and the hole that are located in different Xenes and phosphorene or TMTC monolayers, separated by a
distance $D=h+Nl_{\text{hBN}}$. $l_{\text{hBN}}=0.333$ nm is the thickness of the hBN layer, $h$ is the average thickness of the Xene and phosphorene or TMTC, and $N$ is the number of hBN layers. 

Following the standard procedure~\cite{Landau} for the separation of the
relative motion of the electron-hole pair from their center-of-mass motion,
one introduces variables for the center-of-mass of an electron-hole pair.
After a lengthy
calculation, following Refs. \cite%
{MacDonald1986,Spiridonova,KezSpir2021,RKAS2021b},
one obtains the equation that describes the Mott--Wannier
exciton in the Rydberg optical states in the external electric and  magnetic field
perpendicular to the Xenes/hBN/BP heterostructure. Finally, the equation for the
relative motion of the electron and hole in the Xenes/hBN/BP heterostructure
in the center-of-mass system (cf. Appendix and Refs. \cite%
{Gorkov1967,MacDonald1986,Lozovik1997}) reads
\begin{equation}
\left[ -\frac{1}{2 \mu_x}\frac{\partial ^2}{\partial x^2}-\frac{1}{2\mu_y}%
\frac{\partial ^2}{\partial y^2}+\frac{e^2}{8 \mu_x}B^2x^2+\frac{e^2}{8 \mu_y%
}B^2 y^2+V(x,y)\right] \Phi (x,y)=\mathcal{E}\Phi (x,y).  \label{eq:finsch}
\end{equation}
where
\begin{equation}
\mu_{x} = \frac{m^{e} m_{x}^{h}}{m^{e}+ m_{x}^{h}}   \  \    \text{and} \  \
\mu_{y} = \frac{m^{e} m_{y}^{h}}{m^{e}+ m_{y}^{h}}
\label{eq:ReduceMassAn}
\end{equation}
are the reduced
masses in the $x$
and $y$ directions, respectively. In Eq. (\ref{eq:finsch}) the anisotropy is present in the kinetic and magnetic terms, while the
potential term has isotropic form and the action of the electric field is present via anisotropic reduced masses $\mu_{x}$ and $\mu_{y}$. We solve Eq. (\ref{eq:finsch}) using $V\left( \sqrt{x^{2}+y^{2}+D^{2}}\right)$ for indirect magnetoexcitons. Note that Eq. (\ref{eq:finsch}) does not explicitly contain any spin- or valley-dependent Zeeman terms.

As discussed in the Appendix, the center-of-mass dynamics is described by a harmonic oscillator of mass $M$
and with frequency $\varpi =\frac{eB}{2\sqrt{M\mu}}$ for the isotropic system with $\mu_x=\mu_y$.
Thus, the energy of the interacting electron–hole pair in a 
magnetic field is the sum of the energies of the relative motion and the center-of-mass motion.
Notice that when the exciton is considered beyond the center-of-mass frame, the separation of collective (center-of-mass) and internal motions for two particles in a magnetic field becomes a long-standing and challenging problem~ \cite{Elliott1960,
Gorkov1967,Shinada1965,Akimoto1967,Shinada1970,Avron1978,Lozovik1978,Herold1981,MacDonald1986,Stafford1990,
Lozovik1997}. Such separation is only possible for two particles with equal masses and equal magnitudes of charge. In this special case, the center-of-mass motion becomes quantized and forms the Landau levels.

For excitons in our heterostructures, this separation is not possible. The Landau orbit of the center-of-mass and the internal excitonic motion are strongly coupled~\cite{Schmelcher1991}. In Ref.~\cite{Schmelcher1991}, the authors investigate several physical scenarios in which this coupling becomes significant even at laboratory magnetic-field strengths. Notably, the Landau orbit itself can be substantially modified by this interaction.
In contrast, for Rydberg exciton states, our system does not fall into that category. Therefore, for describing the internal exciton motion, we follow the approaches developed in~\cite%
{Gorkov1967,MacDonald1986,Lozovik1997}. 

The Schr\"{o}dinger equation with
Hamiltonian~(\ref{eq:Schredingermag}) for Xenes/hBN/TMDC heterostructure when for TMDC monolayer $m_{x}^{e}=m_{x}^{h}\equiv m^{h}$ has the form: $\hat{H}\Psi (\mathbf{r}%
_{e},\mathbf{r}_{h})=E\Psi (\mathbf{r}_{e},\mathbf{r}_{h})$, where
$\Psi (\mathbf{r}_{e},\mathbf{r}_{h})$ and $E$ are the
eigenfunction and eigenenergy, respectively. Due to the spherical symmetry, in this case, the corresponding equation easier write in spherical coordinates.
Finally, after separating the angular variable %in (\ref{Relmotion}),
the equation in the center-of-mass momentum system  \cite{MacDonald1986, Lozovik1997}:

\begin{equation}
\left[ -\frac{1}{2\mu}\frac{{\partial }^{2}}{\partial r^{2}}-\frac{1}{2\mu}\frac{1}{r}\frac{{\partial }}{\partial r}+\frac{%
e^{2}}{8 \mu}B^2r^{2}+V(r)\right] \Phi (r)=E\Phi (r),  \label{eq:finsch1}
\end{equation}%
where
\begin{equation}
%\gamma =\frac{m_{h}-m_{e}}{m_{h}+m_{e}}$ and
\mu =\frac{m^{e}m^{h}}{%
m^{e}+m^{h}}
\label{eq:ReduceMassIs}
\end{equation}
is the reduced mass. 
This equation describes the Mott--Wannier magnetoexciton in Rydberg
optical states in 2D materials. 
Here we have neglected the terms that couple the center-of-mass and the relative
coordinates, an approximation that is validated within our calculation.
Eq. (\ref{eq:finsch}) has a long history in the
context of the electron-hole Coulomb interaction \cite{Elliott1960,
Gorkov1967,Shinada1965,Akimoto1967,Shinada1970,Lozovik1978,Herold1981,MacDonald1986,Stafford1990,
Lozovik1997}.

In addition to the exciton binding energy, another important quantity is the diamagnetic coefficient, which gives the change of the exciton binding energy by an external magnetic field:
\begin{equation}
\Delta E = |E(B)-E_0|,
\end{equation}
where $E(B)$ is the binding energy of magnetoexciton, and $E_0$ is the binding energy of a
conventional exciton. The Taylor expansion of $\Delta E$ in powers of $B$ can be used to
determine the diamagnetic coefficient.
Considering the first three terms of the Taylor series \cite{Walck,Godoy2006,Abbarchi2010}
we get
\begin{equation}
E(B) = E_0+\gamma_1 B +\gamma_2 B^2 +O(B^3) \label{eq:Taylor_ser}
,
\end{equation}
where the second and third term are the Zeeman and diamagnetic shifts, respectively.
For Rydberg states our model has only contribution from the diamagnetic shift. The
diamagnetic coefficient is defined as
\cite{Rogers1986,Nash1989,Walck,Erdmann2006,Stier2016, Liu2019,Gor2019}:
\begin{equation}
\gamma_2 = \frac{e^2}{8\mu}\langle r^2 \rangle, \label{eq:dmc}
\end{equation}
where $\langle r^2 \rangle$ is the expectation value of $r^2$ over exciton envelope wave function.
The diamagnetic coefficient quantifies the extent to which a material resists changes in its magnetization when subjected to an external magnetic field. The value of a diamagnetic coefficient depends on the exciton binding energy, because the strength of the exciton binding energy affects how it responds to an external field. The magnetic field strength alters the size of the exciton wavefunction and modifies the energy levels. The dimensionality and the structure of the material affect the
exciton radius and lead to a different diamagnetic response.

\begin{figure}[b]
\begin{centering}
\includegraphics[width=18.0cm]{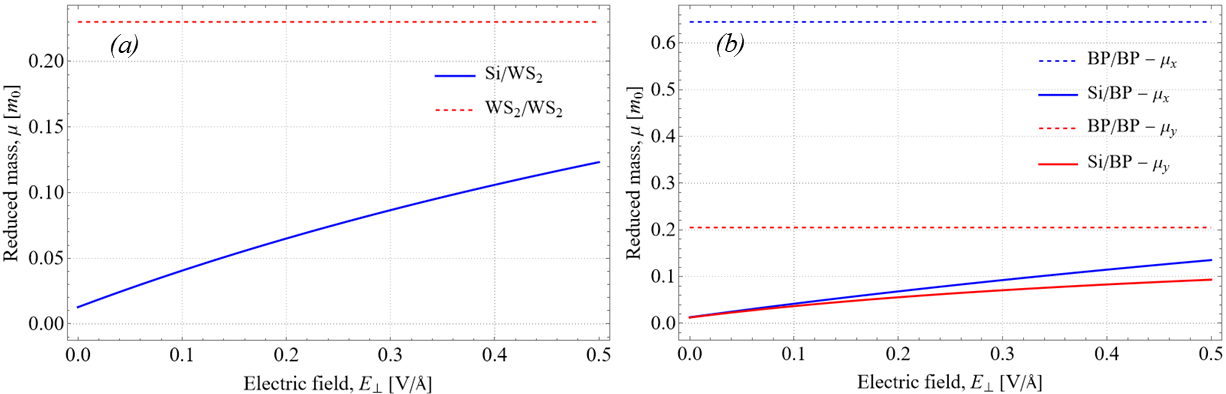}
\caption{(Color online) The dependence of the reduced mass of an exciton in
Xenes/TMDC ($a$) and Xenes/BP ($b$) heterostructures
 on the electric field.  In calculations, we used masses of holes in WS$_{2}$ and BP monolayers from \cite{Kylanpaa2015} and \cite{Peng2014}, respectively. 
}
\label{fig2M}
\end{centering}
\end{figure}

\begin{figure}[t]
\begin{centering}

\includegraphics[width=15cm]{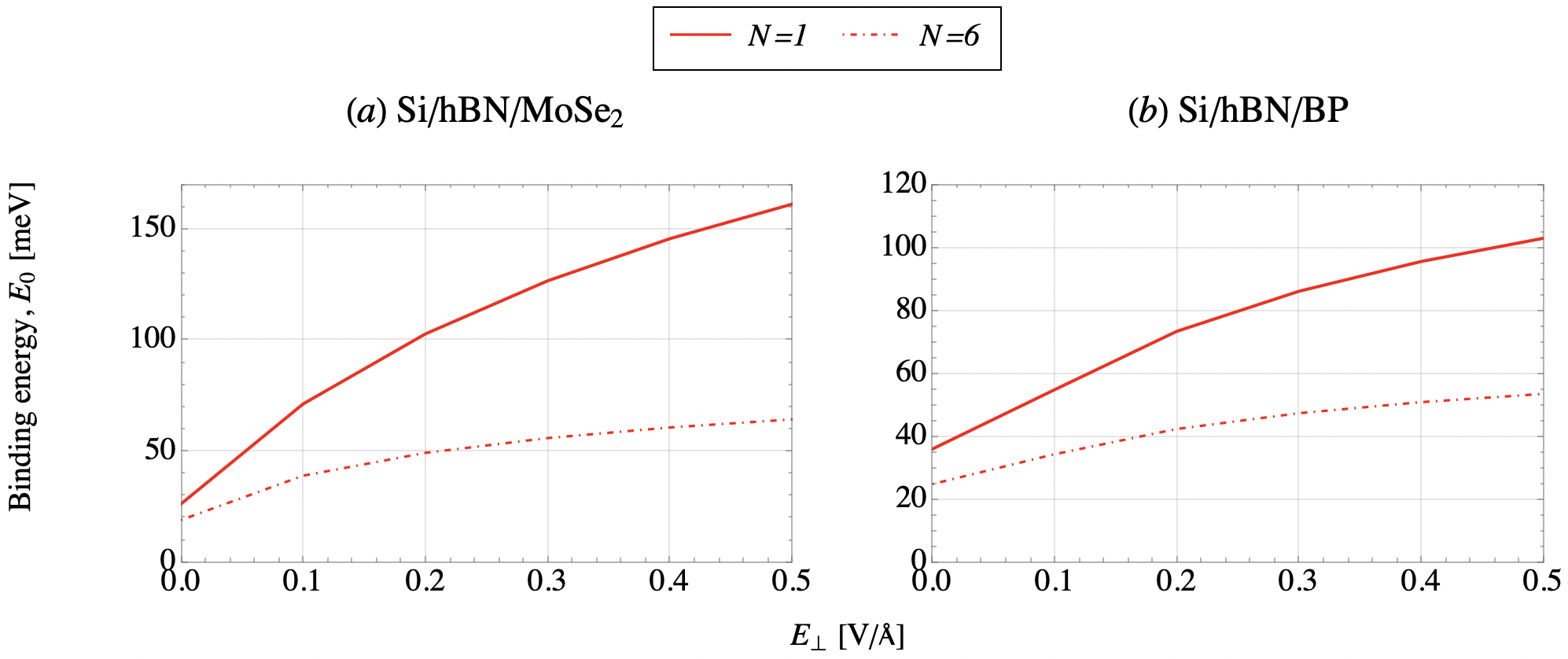}

\caption{(Color online) The electric-field dependence of the binding energies of an
exciton in the Rydberg state 1$s$ for $(a)$ Si/hBN/MoSe$_2$ and $(b)$
Si/hBN/BP heterostructures. The number of dielectric layers, hBN,
is given for $N$ = 1, 6. In calculations, we used masses of holes and thicknesses of MoS$_{2}$ and BP monolayers from \cite{Ramasubramaniam2012} and  \cite{Kylanpaa2015}, and \cite{Peng2014} and \cite{Kumar2016}, respectively.
}
\label{fig3M}
\end{centering}
\end{figure}
\section{Results of Calculations and Discussion}
\label{results}

First, let us focus on the reduced mass in the Xenes/hBN/TMDC and Xenes/hBN/BP heterostructures.
There is interesting physics here, but the presentation can benefit from clarification. The magnetic contribution to the exciton binding energy is inversely proportional to the exciton’s reduced mass, which is always smaller than the smaller of the two constituent masses. In TMDCs and phosphorene, the hole effective mass typically exceeds that of the electron. In contrast, Xenes exhibit equal and electrically tunable electron and hole effective masses ($m^e = m^h$). Consequently, by localizing the hole in BP, TMDC, or TMTC layer, one can achieve a substantial variation in the reduced mass—potentially reaching values smaller than the electron effective mass in TMDCs or BP.

This dependence of the reduced mass on the heterostructure composition is illustrated in Fig. \ref{fig2M}. The analysis reveals that the reduced mass in Xenes/TMDC and Xenes/BP heterostructures is smaller than in TMDC/TMDC or BP/BP counterparts. As follows from Fig. \ref{fig2M}, this reduction of mass enhances the magnetic contribution to the binding energy and increases its tunability with an applied electric field.

For the calculations of reduced masses, we used Eqs.~(\ref{eq:ReduceMassIs}) and~(\ref{eq:ReduceMassAn}) for isotropic and anisotropic heterostructures, respectively. The effective masses of electrons and holes in TMDC materials were obtained using various methods~\cite{Kormanyos2015}. In Xenes/hBN/TMDC heterostructures, we used hole effective mass values from Refs.~\cite{Ramasubramaniam2012,Berkelbach2013,Chernikov2014,Kormanyos2015}, while the electron mass was calculated using Eq.~(\ref{eq:effmassEz}).

In calculations for the Xenes/hBN/BP heterostructure, the effective masses of electrons and holes can be taken from Refs.~\cite{Qiao2014,Peng2014,Tran2014,Paez2016,Kumar2016}, which are based on first-principles calculations. However, the lattice constants reported in these references differ, as do the exchange-correlation functionals and tight-binding parameter sets used in the simulations. These variations naturally lead to discrepancies in the anisotropic effective masses, particularly in the curvatures of the conduction and valence bands along the armchair and zigzag directions. In calculations for Xenes/hBN/TMTC heterostructures, one can utilize the data for effective electrons and holes masses reported in \cite{Jin2015,Torun2018,Donck2018,Wang2020} that are summarized in \cite{KezSpir2022tmtc}. In our calculations, we use input parameters listed in Table \ref{Imputtable}.

\textit{Ab initio} calculations \cite{Drummond2012} indicate that the crystal structure of silicene becomes unstable at electric fields around 2.6 V/\AA. The band gap in Xenes can be tuned by an external electric field up to this critical strength. However, in the proposed van der Waals heterostructures, the Xene layer is separated from the adjacent 2D monolayer by insulating hBN spacers. The electric breakdown field of hBN monolayer is approximately 0.1 V/\AA, although it depends on both the thickness and the number of hBN layers\cite{Lee2011,Britnell2012}. A recent study \cite{Weintrub2022} reports that, under certain conditions, the breakdown threshold for a monolayer can be increased to about 0.4 V/\AA. Accordingly, in our calculations, we considered electric field strengths in the range of 0.1-0.5 V/\AA.

\begin{table}[t]
\caption[]{Input parameters. Reduced masses in units of the mass of electron $m_{0}$ }

\label{Imputtable}
\begin{center}
\begin{tabular}{ccccc}
\hline\hline
\noalign{\smallskip}
\multicolumn{5}{c}{Xenes} \\ \hline
\noalign{\smallskip}
& 2$\Delta _{so}$ [meV] & $d_{0}$ [\AA] & $v_{F}\times 10^{5}$
[m/s] & $h$ [\AA] \\ \hline
\noalign{\smallskip}
Si & 38 \cite{Li2013} & 0.46 \cite{Li2013} & 5.06 \cite{Li2013} & 3.33 \cite{Li2013}\\ \hline
\noalign{\smallskip}
\multicolumn{5}{c}{BP and TMTC} \\ \cline{2-5}
\noalign{\smallskip}
&  & $\mu _{x}^{e}$ [$m_{0}$] & $\mu _{y}^{h}$ [$m_{0}$] & $h$ [\AA] \\ 
\cline{2-5}
\noalign{\smallskip}
& BP &  0.64496 \cite{Peng2014} & 0.20488 \cite{Peng2014} &   5.41 \cite{Kumar2016} \\ 
& ZrS$_3$ & 1.26 \cite{Jin2015} &  0.33  \cite{Jin2015}&  8.9  \cite{Jin2015}\\ \cline{2-5}
\noalign{\smallskip}
\multicolumn{5}{c}{TMDC} \\ \cline{2-4}
\noalign{\smallskip}
&  & $\mu $ [$m_{0}$] & $h$ [\AA] &  \\ \cline{2-4}
\noalign{\smallskip}
& WS$_{2}$ &  0.23 \cite{Ramasubramaniam2012} & 6.219 \cite{Kylanpaa2015}   \\ 
& MoSe$_{2}$ &  0.31 \cite{Ramasubramaniam2012} & 6.527 \cite{Kylanpaa2015}   \\ \cline{2-4}
\end{tabular}
\end{center}
\end{table}

 We numerically solve Eqs. (\ref{eq:finsch}) and (\ref{eq:finsch1}) using a code implemented in Refs. \cite{KezSpir2022tmtc,KezSpir2022phosp,Spiridonova,KezSpir2021,RKAS2021b}. The method utilizes the finite element method implemented in Wolfram Mathematica in the NDEigensystem function, which yields pairs of eigenenergies and eigenfunctions corresponding to the most strongly bound states. Parameters are taken from above above-mentioned references and summarized in Table \ref{Imputtable}. Notice that parameters are not as important since in the literature for each material, there is a range for each parameter. Therefore, our work demonstrates the viability of the systems rather than exact numbers. The results change based on the parameters used.

As an illustrative example, we present below the results for $A$-type indirect excitons. Including $B$-type excitons does not lead to any qualitative changes in the conclusions.

The relationship between the binding energy (BE) and the reduced mass of an exciton arises from the nature of the Coulomb interaction between the electron and the hole. The exciton binding energy, defined as the energy required to dissociate the exciton into free electron and hole, is proportional to the reduced mass of the electron–hole system. Thus, a larger reduced mass leads to a stronger binding energy. Consequently, as the electric field increases, the reduced mass also increases, resulting in a corresponding increase in the BE. This dependence is illustrated in Fig.~\ref{fig3M} for excitons in the Rydberg state 1$s$. Also, from Fig.~\ref{fig3M}, one can conclude that increasing the number of hBN layers results in a notable reduction in the exciton binding energy due to enhanced dielectric screening effects.

The relationship between BE and reduced mass has significant implications in monolayer physics. In TMDC/hBN/TMDC heterostructures, which exhibit higher reduced masses than
Xenes/hBN/TMDC, excitons tend to have higher binding energies. This characteristic
behavior
affects the efficiency of exciton-based devices, as lower binding energies facilitate the dissociation of excitons into free charge carriers—an essential process for efficient charge
transport. Understanding this dependence is crucial for the design of materials with tailored
electronic and optical properties, particularly in systems where excitonic effects play a dominant role.

\begin{figure}[t]
\begin{centering}

\includegraphics[width=8.5cm]{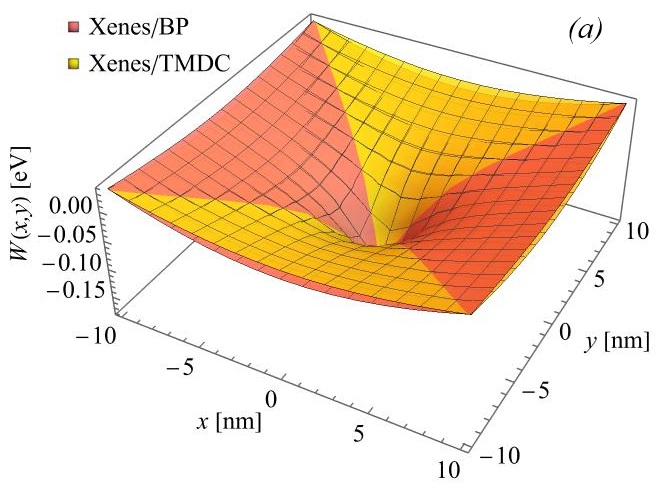}
\includegraphics[width=8.5cm]{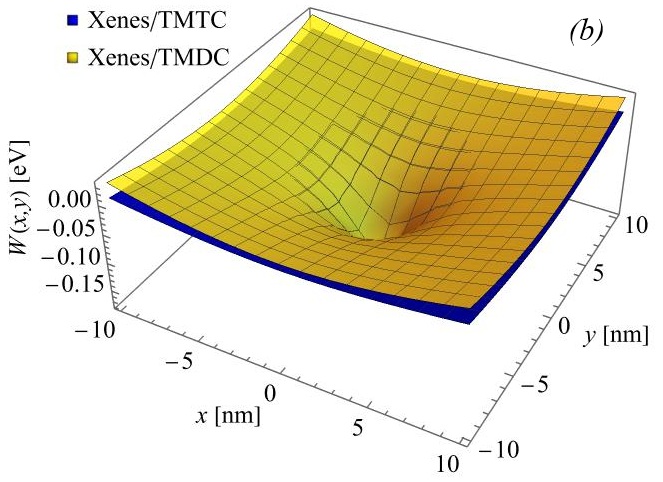}
\caption{(Color online) The dependence of the total potentials $W_{1}(x,y)=\frac{e^{2}}{8\mu _{x}}B^{2}x^{2}+\frac{%
e^{2}}{8\mu _{y}}B^{2}y^{2}+V(x,y)$ and $W_{2}(x,y)=\frac{e^{2}}{8\mu }B^{2}(x^{2}+y^{2})+V(x,y)$ acting on the electron-hole pair in Xenes/BP (Si/BP) and Xenes/TMDC (Si/WS$_{2}$) ($a$) and  Xenes/TMTC (Si/ZrS$_{3}$) and Xenes/TMDC (Si/WS$_{2}$) $b$) heterostructures. Calculations are performed for the electron-hole pair in the external magnetic field $B=30$ T and electric field $E_{\perp}=0.1$ V/\AA.}
\label{Potentials}
\end{centering}
\end{figure}

\begin{figure}[h]
\begin{centering}
\includegraphics[width=15.0cm]{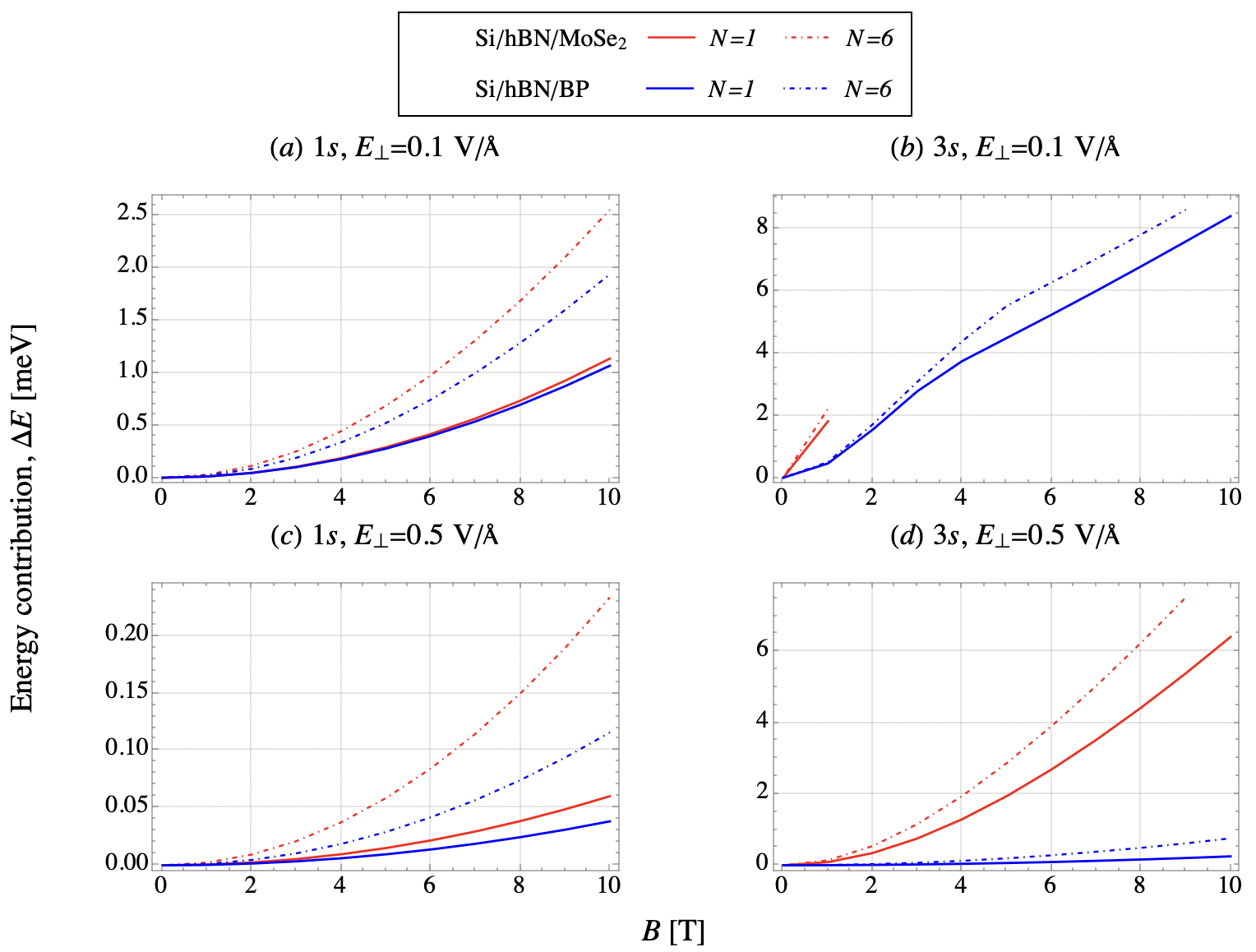}
\caption{(Color online) Dependence of energy $\Delta E$ for magnetoexcitons
in $1s$ and $3s$ states in Xenes/MoSe$_2$ and Xenes/hBN/BP heterostructures on the electric and magnetic fields.  In calculations, we used parameters summarized in Table \ref{Imputtable}.
}
\label{comparison}
\end{centering}
\end{figure}

In the next step, let us consider indirect magnetoexcitons formed by electron-hole pair in van der Waals heterostructures (e.g., Xenes/hBN/TMDC, Xenes/hBN/BP, Xenes/hBN/TMTC), in the presence of a magnetic field. In these vdW heterostructures magnetoexcitons emerge as a result of the influence of an external magnetic field on excitons, \textit{i.e.} the presence of an external magnetic field alters the excitonic states and leads to the formation of magnetoexcitons, which are excitons whose properties are modified by the magnetic field. The characteristics of magnetoexcitons are governed by both the underlying electronic structure of the heterostructure and the strength of the magnetic field, modifying their internal structure and energy spectrum. In Eqs. (\ref{eq:finsch}) and (\ref{eq:finsch1}) $W_{1}(x,y)=\frac{e^{2}}{8\mu _{x}}B^{2}x^{2}+\frac{e^{2}}{8\mu _{y}}B^{2}y^{2}+V(x,y)$ and $W_{2}(x,y)=\frac{e^{2}}{8\mu }B^{2}(x^{2}+y^{2})+V(x,y)$ are total potentials that act on the
electron-hole pair in these heterostructures, respectively. The Coulomb interaction $V(x,y)$ (\ref{eq:Schredingermag}) as well as $W_{2}(x,y)$ have a rotational symmetry.
In contrast, though, the terms $\frac{e^2}{8\mu_x}B^2 x^2	$ and $\frac{e^2}{8 \mu_y}B^{2}y^{2}$ in $W_1 (x,y)$ break the rotational symmetry due to its anisotropic nature.
 As a result of reduced masses anisotropy, when $\mu_{x}>\mu_{y}$, $\frac{e^{2}}{8\mu _{y}}B^{2}y^{2}$ gives larger
contribution due to the magnetic field than the term $\frac{e^{2}}{8\mu _{x}}
B^{2}x^{2}$ and vice versa if $\mu _{y}>\mu _{x}$. Thus, the electron and hole masses anisotropy leads to the anisotropy of $W_{1}(x,y)$. The dependence of the total potential $W_{1}(x,y)$ and $W_{2}(x,y)$, respectively, acting on the electron-hole pair on $x$
and $y$ coordinates is shown in Fig. \ref{Potentials}. Let us emphasize that the anisotropy of the potential $W_{1}(x,y)$ with respect to $W_{2}(x,y)$ is different in the case of heterostructures with phosphorene and TMTC.

The results of calculations of the energy contribution $\Delta E$ due to a
magnetic field at different values of the electric field are presented in
Fig. \ref{comparison}. Here, we consider MoSe$_2$ and phosphorene as representative
cases for isotropic and anisotropic materials, respectively.
Contributions of the magnetic field are proportional to the factors $1/\mu$ for an isotropic material and $1/\mu_{x}$ and $1/\mu_{y}$ for an anisotropic material. Therefore, the electric field increase leads to the decrease of $1/\mu$, $1/\mu_{x}$,
and $1/\mu_{y}$, consequently, $\Delta E$ decreases.  At the small electric field $1/\mu$, $1/\mu_{x}$, and $1/\mu_{y}$ are comparable, although $1/\mu_{x}$, and $1/\mu_{y}$ a little bit exceed $1/\mu$.
By increasing the number of hBN layers, the electrostatic attraction decreases,
which changes $\Delta E$. In Fig. \ref{comparison},  we demonstrate this for $1s$ and $3s$ states, but the same tendencies can be extended to $2s$ and $4s$ states.
%When $N=1$ the difference between $\Delta E$ for both heterostructures is negligible (Fig. \ref{comparison} $a$ - $b$). When $E_{\perp} >1$ V/m, $1/\mu < 1/\mu_{x}$ and $1/\mu < 1/\mu_{y}$, and $\Delta E$ becomes larger for heterostructure with phosphorene (Fig. \ref{comparison} $c$ - $f$). In Fig. \ref{comparison},  we demonstrate this for $1s$ and $3s$ states, but the same tendencies can be extended to $2s$ and $4s$ states.
In Xenes/hBN/MoSe$_{2}$ and Xenes/hBN/BP heterostructures, the strong field-tunable band gaps and anisotropic effective masses of phosphorene can lead to pronounced shifts in exciton binding energy and modified magnetoexciton behavior compared to conventional heterostructures built with TMDC or phosphorene monolayers. Due to the larger effective dielectric constant introduced by hBN layers, excitons are more weakly bound, making their magneto-optical response more sensitive to external fields.

The analysis of these calculations leads to the following conclusions: (i) the energy contribution increases due to the increase of magnetic field; (ii) The addition of hBN layers gives an increase in $\Delta E$ energy; (iii) The increase in the electric field leads to the decrease of $\Delta E$.

The diamagnetic coefficients for excitons in Si/hBN/MoSe$_2$ and Si/hBN/BP heterostructures, obtained in the framework of our approach, are reported in Table \ref{table1}. For Xenes/hBN/TMDC we show a representative case - Si/hBN/MoSe$_2$. Analysis of the presented results shows: i. the diamagnetic coefficients are decreasing with increasing electric field for all Rydberg states in both heterostructures; ii. the increase in the number of insulating hBN layers leads to the increase of $\gamma_2$ for $1s$, $2s$ and $3s$ states; iii. $3s$ and $4s$ states generally exhibit high
diamagnetic coefficients due to the increase exciton's $r^2$ affected by the size of the exciton wavefunction.

\begin{figure}[t]
\begin{centering}
\includegraphics[width=8.0cm]{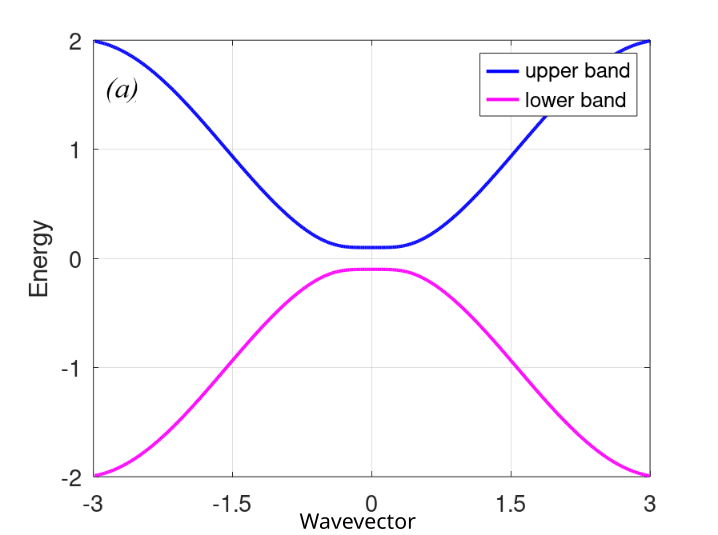}
\includegraphics[width=8.0cm]{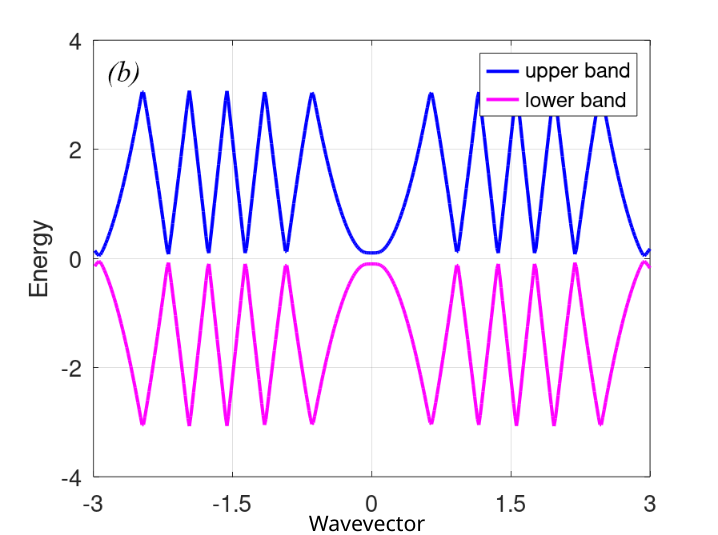}
\caption{(Color online) Typical effect of Floquet band engineering: while ($a$) represents the
static case, ($b$) is the result of the Floquet dynamics due to a periodically
time-dependent exciton mass, which represents a backfolded spectrum on the interval
$[-\pi,\pi]$.  Since the Hamiltonian (\ref{HamPer}) is isotropic, the direction of the wavevector does not affect the spectrum.
}
\label{fig:floquet}
\end{centering}
\end{figure}

\begin{table}[t]
\caption[]{Diamagnetic coefficients for magnetoexcitons in Si/hBN/BP and Si/hBN/MoSe$_2$. DMCs are %given in $\mu$eV/T$^2$ and
obtained when $R^2 = 0.9998$ for the linear regression model. Some cells are empty because diamagnetic coefficients could not be extracted with linear regression satisfying $R^2 = 0.9998$. 
}
\label{table1}
\begin{center}

\begin{tabular}{ccc|cc}
\hline\hline

& \multicolumn{4}{c}{Diamagnetic coefficient $\gamma_2 $ [$\mu $eV/T$^{2}$]}
\\ \hline
& \multicolumn{2}{c}{Si/hNB/BP} & \multicolumn{2}{|c}{Si/hBN/MoSe$_{2}$} \\
\cline{2-5}

\multicolumn{5}{c}{$N=1$} \\ \hline

$E_{\perp }$ [V/\AA] & 0.1 & 0.5 & 0.1 & 0.5 \\ \hline

1$s$ & 10.70 & 1.18 & 11.36 & 0.60   \\
2$s$ & & 4.38 &  & 13.89  \\  \hline

\multicolumn{5}{c}{$N=6$} \\ \hline

1$s$ & & 3.05 & 25.66 & 2.34  \\

2$s$ & &   & & 26.59  \\ \hline\hline
\end{tabular}
\end{center}
\end{table}
Our results suggest the consideration of a time-periodic electric field with period
$T$ that,
according to our single-exciton calculation, creates a time-periodic exciton mass.
Then the resulting time-periodic Hamiltonian $H_{\bf p}(t)$ can be treated within the Floquet
approach~\cite{Lindner2011}. This yields a Floquet Hamiltonian, which
describes a single exciton. However, a more realistic situation is to consider a condensate of excitons.
Having calculated the properties of a single exciton in external fields,
we can extend this case to an excitonic condensate in the double-layer structure. Assuming
that the excitons have a small dipolar moment and interact weakly with each other, we can
ignore the exciton-exciton interaction in the first order approximation and
calculate the band structure of the non-interacting excitonic gas. To maintain translational
invariance, we simplify the situation further and consider the case without an external
magnetic field.
This leads to a translation-invariant distribution of excitons, which we describe in
Fourier representation with momentum ${\bf p}$ as the effective
Hamiltonian~\cite {LozovikYudson}, which was mentioned at the beginning of
Sect. \ref{sect:model}:
\begin{equation}
H_{\bf p}(t)=\begin{pmatrix}
p^2/2\mu(t) & \Delta_p(t) \\
\Delta_p(t) & -p^2/2\mu(t)
\end{pmatrix}
,
\label{HamPer}
\end{equation}
where $\Delta_p(t)$ represents the band splitting due to electron-hole pairing
and $\mu(t)$ the time-dependent reduced mass. $\Delta_p(t)$ can be calculated self-consistently,
following the approach of Ref.~\cite {LozovikYudson}, where the mass oscillations are
corrections in comparison to the time-average mass. At low density, $\Delta_p>0$ for a sufficiently
strong interlayer attraction between electrons and holes. The other condition is that
the gap $2\Delta_{p\xi\sigma}=2|\xi \sigma \Delta _{so}-ed_{0}E_{0}\cos wt|$
must be larger than the kinetic energy $p^2/2\mu$.
This means that $E_0$ must be smaller than $\Delta_{so}/ed_0$.
%And (ii) $\Delta_p$ must be smaller than the excitonic binding energy.
%$\Delta_{p\xi\sigma}=|\xi \sigma \Delta _{so}-ed_{0}E_{0}\cos wt|$ with
%$0<E_{0}< |\xi \sigma \Delta _{so}/ed_{0}|$
Thus, we can assume within the leading approximation that only the reduced mass $\mu(t)$ depends on
the external electric field. From $H_{\bf p}(t)$ we obtain the Floquet Hamiltonian~\cite{Lindner2011}
as
\begin{equation}
H_{F;{\bf p}} = \frac{i}{T}\log\left[\prod_{t=0}^{T}e^{-i{\tilde H}_{\bf p}(t)dt}\right]
.
\end{equation}
Its eigenvalue spectrum yields the effective band structure of the Floquet approach.
This can be understood in the present case as if the variation of the reduced mass
$\mu(t)$ increases and decreases the bandwidth periodically in time. Since the Floquet
Hamiltonian is created by the unitary operator, values of the spectrum of
$H_{\bf p}(t)$ that are outside the interval $[-\pi,\pi]$ are backfolded on
the complex unit circle. This leads to the restriction of the bandwidth to $2\pi$.
This effect is illustrated in Fig. \ref{fig:floquet}, which reveals that the time-dependent
electric field applied to the layered system provides a tool for band-structure
engineering, similar to the concept presented in Ref.~\cite{Zhan2024}.

 \section{Concluding remarks}
 \label{ConcludRemark}
In this work we suggest the novel vdW heterostructures, comprising Xenes, TMDCs, phosphorene,
and TMTCs monolayers separated by hBN insulating layers. We have investigated theoretically
the properties of Rydberg indirect excitons in Xenes/hBN/TMDC, Xenes/hBN/BP, and Xenes/hBN/TMTC
heterostructures subjected to parallel external electric and magnetic fields that are oriented
perpendicular to the layers. By incorporating both isotropic and anisotropic materials, we
demonstrated the ability to control and tune excitonic properties through external fields and
structural parameters.

Our study highlights several key findings: (i) the exciton reduced mass and, consequently, the binding energy can be effectively modulated by the strength of the electric field and the number of hBN layers; (ii) anisotropy in the effective mass leads to distinguishable trends in binding energy and diamagnetic response when compared to isotropic systems; and (iii) the energy contribution from the magnetic field and the diamagnetic coefficients decrease with increasing electric field, while increasing the number of hBN layers enhances these quantities due to reduced Coulomb interaction.

Furthermore, we explored the application of a time-periodic electric field as a potential method for band structure engineering. This opens avenues for dynamic control of excitonic states in 2D material systems, paving the way for novel optoelectronic and quantum devices based on excitonic phenomena.

In summary, vdW heterostructures, composed of two layers of 2D materials when one of the layers
is Xenes, offer a versatile platform for exploring novel electronic and optical properties.
By carefully engineering the stacking, orientation, and strain of these materials and applying
external electric and magnetic fields,  researchers can design systems with tailored
properties. %, which are suitable for a wide range of applications in next-generation electronic and optoelectronic devices.

\appendix
\section{Approximation to decouple the relative and center-of-mass motions}
\renewcommand{\theequation}{A\arabic{equation}}
\setcounter{equation}{0}

The Schr\"{o}dinger equation of an interacting $e$–$h$ pair in a magnetic field $\mathbf{B}$ reads
\begin{equation}
\left[ \frac{1}{2m^{e}}\left( \mathbf{p}_{e}-e\mathbf{A}_{e}\right)^{2} 
+ \frac{1}{2m^{h}}\left( \mathbf{p}_{h}+e\mathbf{A}_{h}\right)^{2} 
+ V(|\mathbf{r}_{e}-\mathbf{r}_{h}|)\right]
\Psi(\mathbf{r}_{e},\mathbf{r}_{h})=\mathcal{E}\,\Psi(\mathbf{r}_{e},\mathbf{r}_{h}),
\label{Magexiton}
\end{equation}
where $\mathbf{p}_{e}=-i\hbar \nabla_{e}$ and $\mathbf{p}_{h}=-i\hbar \nabla_{h}$, %$m^{e}$ and $m^{h}$ are the effective masses of the electron and hole, 
$\mathbf{A}_{i}=-\mathbf{r}_{i}\times \mathbf{B}/2$ is the vector potential generating the magnetic field $\mathbf{B}$, 
and $V(|\mathbf{r}_{e}-\mathbf{r}_{h}|)$ is an electrostatic interaction. 
We assume the symmetric gauge corresponding to a uniform perpendicular external magnetic field $\mathbf{B}=(0,0,B)$. 

Our focus is on Rydberg $s$-state excitons. 
The terms $e\mathbf{p}_{e}\mathbf{A}_{e}$ and $e\mathbf{p}_{h}\mathbf{A}_{h}$ in Eq.~(\ref{Magexiton}) 
lead to an effective electric field that appears in the center-of-mass frame of the $e$–$h$ system which interacts with the relative coordinate electric
dipole $er$, and orbital momentum of the $e-h$. However, orbital momentum of the $e-h$ pair for Rydberg excitons is zero. 
Since these terms are proportional to the magnetic field $B$, restricting ourselves to contributions proportional to $B^{2}$, 
we obtain
\begin{equation}
\left[ \frac{1}{2m^{e}}\left( \mathbf{p}^{2}_{e}+e^{2}\mathbf{A}_{e}^{2}\right)
+ \frac{1}{2m^{h}}\left( \mathbf{p}^{2}_{h}+e^{2}\mathbf{A}_{h}^{2}\right)
+ V_{ij}(|\mathbf{r}_{i}-\mathbf{r}_{j}|)\right]
\Psi(\mathbf{r}_{e},\mathbf{r}_{h})=\mathcal{E}\,\Psi(\mathbf{r}_{e},\mathbf{r}_{h}).
\label{MagTrion}
\end{equation}
At the next step
%, following the standard procedure, 
we introduce the coordinates of the center-of-mass and relative motion for two particles. 
Making the assumption that the electron and hole have nearly equal masses, we obtain
\begin{eqnarray}
\bigg[ -\frac{1}{2M}\frac{\partial^{2}}{\partial \mathbf{R}^{2}}
-\frac{1}{2\mu}\frac{\partial^{2}}{\partial \mathbf{r}^{2}}
+\frac{e^{2}}{8\mu}(\mathbf{B}\times \mathbf{R})^{2}
+\frac{e^{2}}{8\mu}(\mathbf{B}\times \mathbf{r})^{2}
+ V(\mathbf{r}) \bigg] \psi(\mathbf{R},\mathbf{r})
=\mathcal{E}\,\psi(\mathbf{R},\mathbf{r}),
\label{eq:HRrex}
\end{eqnarray}
where $M=m^{e}+m^{h}$ is the total mass and $\mathbf{R}=(X,Y)$.  

Finally, seeking the solution of Eq.~(\ref{eq:HRrex}) in the form $\psi(\mathbf{R},\mathbf{r})=\Psi(\mathbf{R})\Phi(\mathbf{r})$, 
and separating the angular variable, the problem decouples into two equations describing the relative and 
the center-of-mass motion:
\begin{equation}
\left[ -\frac{1}{2\mu}\frac{\partial^{2}}{\partial r^{2}}
-\frac{1}{2\mu}\frac{1}{r}\frac{\partial}{\partial r}
+\frac{e^{2}}{8\mu}(\mathbf{B}\times \mathbf{r})^{2}
+ V(r)\right] \Phi(r)=E\Phi(r),
\end{equation}
\begin{equation}
\left[ -\frac{1}{2M}\frac{\partial^{2}}{\partial \mathbf{R}^{2}}
+\frac{e^{2}}{8\mu}(\mathbf{B}\times \mathbf{R})^{2} \right]\Psi_{cm}(\mathbf{R})
=E_{cm}\Psi_{cm}(\mathbf{R}),
\label{eq:finschCM}
\end{equation}
where $E$ and $E_{cm}$ are the eigenenergies of the relative and center-of-mass motions, respectively, 
so that $\mathcal{E}=E+E_{cm}$.  

The Schr\"{o}dinger equation for the center-of-mass (\ref{eq:finschCM}) formally corresponds to the isotropic harmonic oscillator  
with the effective mass $2\sqrt{M\mu}$ in 2D space, oscillated with frequency $\varpi =\frac{%
eB}{2\sqrt{M\mu}}$.
In Cartesian coordinates it can be solved exactly, with the energy spectrum
\begin{equation}
E_{cm}=\hbar \varpi \left(\mathfrak{n}_{X}+\mathfrak{n}_{Y}+1\right)
=\hbar \varpi \left(\mathfrak{N}+1\right),
\label{Ec.m.}
\end{equation}
where $\mathfrak{N}=\mathfrak{n}_{X}+\mathfrak{n}_{Y}$, 
with $\mathfrak{n}_{X}=0,1,2,3,\ldots$ and $\mathfrak{n}_{Y}=0,1,2,3,\ldots$.  
%The energy (\ref{Ec.m.}) will contribute to the total energy of the magnetoexciton.

Thus, within the above-mentioned approximations, we can decouple the relative and the center-of-mass motions. The energy (\ref{Ec.m.}) will contribute to the total energy of the magnetoexciton.

\end{document}